# Simulation of Obstruction Avoidance Generously Mobility (OAGM) Model using Graph-theory Technique

V. Vasanthi and M. Hemalatha
Department of Computer Science, Karpagam University, Coimbatore, India

**Abstract:** An Obstruction Avoidance Generously Mobility (OAGM) model has been introduced for controlling ad-hoc sensor networks and thereby operating emerging fields like military and healthcare services. According to this model, the ability to send a message to a group of users simultaneously, based solely on their geographic location, is desirable by using Mission Critical Mobility model that assumes the obstacle shapes like rectangle or square in the simulation terrain. The OAGM model is developed by grasping the critical situations of military and healthcare services by incorporating the node movement model, hierarchical node organization, placement of obstacle that affect the movement of nodes and also signal propagation. Graph theory technique is used to find the shortest path of the node movement process. The varying number of parameter sets with DSR protocol is analyzed for MCM and OAGM mobility model. The results show OAGM performance is better than MCM.

## INTRODUCTION

Ad-hoc sensor network is an emerging interesting topic in network communication and in particular the ad-hoc sensor network as a research topic. It is a wireless sensing and processing that communicate through co-operation of sensor nodes, with current advances in technology and wireless networks are increasing in popularity. Wireless networks allow users the freedom to travel from one location to another without interruption of their computing services. However, wireless networks require the existence of a wired Base Station (BS) in order to send/receive messages through wireless user. Ad-hoc sensor networks, a subset of wireless networks, allow the formation of a wireless network without the need for a BS. All participating users are in an ad-hoc sensor network agree to accept and forward messages, to and from each other. With this flexibility, wireless ad-hoc networks have the ability to form anywhere, at any time, as long as two or more wireless users are willing to communicate.

Mobility model is a tool has an aspect that can be used easily and in the simplified manner. The following aspects should be there for mobility model:

- It should be easily accessible and it should be sufficiently used in detail
- Easy to understand, how it follows the method and having parameter in an understandable
- Involves better mathematical properties
- It has a clear scope of validity

The main progress of new model is based on low cost and also for Reliability. Using this mobility model great long-term economic and ability to transfer our lives in a mechanic and it gives new technology which is useful and adopted for optimized problems to get new solutions. Mission Critical Networking has recently communicated in environments like natural or man-made disasters and emergency health care services.

In this study, we present an Obstruction Avoidance Generously Mobility model (OAGM) for Ad-hoc and Sensor network. The proposed model can simulate the movement pattern in an emergency and healthcare service where ad-hoc and sensor network is deployed. In order to capture the real mobility characteristics of the nodes taken by humans in obstacle environment, we incorporate hybrid BELLMAN-FORD-DIJSKTRA algorithm in graph theory technique in our model. In the OAGM model, the node chooses the optimal path in a geographic restricted area. It is the shortest path in single source cheapest path with negative cost edges. The obstacle affects the node movements as well as the signal propagation; we incorporate the Voronoi approach for this model.

To evaluate our model, we compare it to another model (i.e.,) Mission Critical Mobility model by a thorough simulation study. The result shows that OAGM model shows the high Packet Delivery Ratio (PDR) and low Delay packets with a short period of time. For evaluating the MCM and OAGM model the Simulation Terrain assigns one rectangle in the center

**Corresponding Author:** M. Hemalatha, Department of Computer Science, Karpagam University, Coimbatore, India

and set the parameter that set varies with the number of nodes and Speed to find the analysis.

## LITERATURE REVIEW

There are several mobility models developed and proposed to provide the research community with a solution for realistic environment on Ad-hoc and sensor networks. Recent surveys summarized the related works in the same field (Johnson *et al*., 2001; Bai and Ahmed, 2004; Bai *et al*., 2003; Vasanthi and Hemalatha, 2012). The above mentioned mobility models are means of assumptions of the free-space area. This assumption will not be useful for realistic environment where geographic restriction models are proposed. Some of realistic models like for indoor and outdoor environments with obstacles as an integral part of many scenarios are operated. So that obstacle mobility model is proposed in the literature (Jardosh *et al*., 2003).

The obstacle mobility model has introduced not only a movement constraint but also it deals with propagation like signal impairments due to the presence of obstacles. It is created for real environments like campus building. In this model the nodes should follow the predefined path which is connected to a limited point in the network areas. By using Voronoi diagram the node positions are split and placed in the simulation terrain. The other model in geographic restriction is pathway model (Ahmed *et al*., 2008) nodes are allowed to move along the path with edges that represent streets and pathways. In context, the city section mobility model (Shohrab and Mohammed, 2009) is appropriate for simulating mobility in the street network of a city and it includes safe driving characteristics. In the environment mobility model the area is divided into geometric and non-geometric area with different mobility factors (Gang *et al*., 2006).

The authors in Christos *et al*. (2012) present an obstacle aware mobility model which is based on the concept of anchors, to define the trajectories in and around the obstacles. It is based on inclusion of obstacles but it has not deals with any special properties of ad-hoc network deployed in Mission Critical Mobility. In Jonahing (2005) incorporates a model for spatial and temporal dependency especially urban environment. It is applied to community based scenarios without focusing on physical impairment of simulation area.

In obstacle based on social networks by using the theory of social networks is applied in patterns with the presence of obstacles. Johansson *et al*. (2009) develop three 'realistic' mobility scenarios to depict the movement of mobile users in real life, including conference scenarios, event coverage and disaster relief scenarios.

In Mission Critical Mobility (Papageorgiou *et al*., 2009) takes obstacles into account in a more generic fashion. The nodes are moving in the entire free space area without any restriction like moving nodes only in the predefined path. In this model they are using edge detection in every time and passing through edge till the destination is reached. It is using the group movement with the group leader. It is using a free move around the area and edge moving for passing the obstacle which obstructed and unobstructed movements. The main disadvantage of this mobility is that, the time taken to reach the destination is more comparable with our implementation (Vasanthi and Hemalatha, 2012). The nodes are mainly controlled or carried by humans in case of emergency. Frameworks are produced in NS-2 (Fall and Varadhan, 1999). Compatible trace file and it is not integrated with simulation core. There are many different simulation frameworks like Trails (Chatzigiannakis *et al*., 2008), Momose (Boschi *et al*., 2008), Mobisim (Mousavi *et al*., 2007). Here we are using NS2 for implementation.

**An overview of geographic restriction mobility models:** In this section we are going to examine the limitation of Random model, the unconstrained motion of nodes. In Random models the nodes are allowed to move freely based on a random position in the simulation field may be 1000*1000, 700*700, 850*850 in any form. In real life applications we have noticed that node movement is important in the environment. The node movement in Realistic Environment is blocked by buildings and other obstacles. Therefore, the nodes may move in pseudo-random way on predefined pathways in the field. These types of characteristics are addressed by path and obstacles are called in this kind of mobility model with geographic restrictions.

There are three main mobility models come under this category:

- **Obstacle mobility model:** In this model they introduce a part of movement in pre-defined path which is connected to a small network (i.e.,) campus like environment (Jardosh *et al*., 2003). By use of the Voronoi diagram the path and points are located and by using this construction the obstacles are placed in the simulation terrain and defined pre-defined path.
- **Pathway mobility model:** In this model the nodes are allowed to move in the predefined edges that represents streets and pathways.
- **Freeway mobility model:** In this model the behavior of vehicles traveling on a freeway. The

movement of a node is restricted to a lane of a freeway (Shohrab and Mohammed, 2009) and is temporally dependent on the previous speed and other vehicles traveling in front on the same lane. Relationship between the speeds at subsequent time slots in periodically.

- **Mission critical mobility model:** It extends the Obstacle Mobility Model by enabling the nodes to select among the points of the area and move around the obstacles in a natural way. It is suited for far wider of scenarios than the OM model. In this model the time taken to find the edge of obstructed obstacle and unobstructed obstacle is more by evaluation (Papageorgiou *et al.*, 2009; Papageorgiou and Varvarigos, 2008; Christos *et al.*, 2012).

Here we are considering only Mission Critical Mobility model which is the only realistic mobility model.

## MISSION MOBILITY MODEL

The main aim of this model is prepared for real life situations like natural disasters, military activities and emergency health care services. In this model the nodes are carried by humans. Each node of this model moves towards the destination freely with minimum value and maximum value. In this model the destination node is selected by uniformly distributed, which can be switched to any type of reflecting the characteristics of a specific scenario, without altering the model itself.

**Node movement:** In this model a destination point is set, each node moves its way around the obstacles following a recursive process in order to reach it. If there is an unobstructed line of sight connecting the node with the destination point, the node follows this direct line to get to the desired destination. If there is an obstacle in the way, the node sets as its next intermediate destination the vertex of the directly visible obstacle edge that is closest to the destination and repeats the same process all over again with starting point i.e., its initial position and destination the chosen vertex. This is repeated until an unobstructed direct line until the current destination is found. The whole process is executed recursively until the destination is reached. This algorithm is explained in the above process for node movement.

The node acts as a group and each one has its own activity mode that is emergency workers and medical staff are included in MCM model. The emergency workers may be like police, fire and soldiers like this urgent activity one completes and another follows. In medical staff type the medics and bearer complete one and return to certain base station BP before going to new events or work (Jardosh *et al.*, 2006). In this section, we have discussed the geographic restriction mobility models and considering the movement of nodes in the simulation environment in MCM. In the real world, the mobile nodes using a Critical Mission Mobility Model move freely in the simulation area (Christos *et al.*, 2012).

We have implemented the Mission Critical Model (MCM) in NS2.33 with the parameter metrics with one obstacle as a one scenario with the variation of speed (between 0 to 10 with the time intervals 2 m/s) and nodes (50 to 250 interval nodes of 50). We observed that the MCM works within the network area. For example a node is located at points S and D as destination point. The obstacle is in between the S and D with using blocking the straight or direct line movement from S and D. The node passes through the vertex and edge and it repeats until it reaches the destination. The time taken for checking the obstacle's edge and the straight line from the new edge until it reaches the destination.

## OBSTRUCTION AVOIDANCE GENEROUSLY MOBILITY (OAGM) MODEL

The main objective of Obstruction Avoidance Generously Mobility (OAGM) model is to realistically stimulate Manet and Sensor Network that can be operated in emergency situations according to human behavior. HYBRID BELLMAN-FORD-DIJKSTRA algorithm for finding shortest path have used in node movement OAGM model, which depends according to the environment. The source and destination nodes are selected randomly anywhere from the simulation area by using the recursive function. The obstacles present in the path of the source and destination nodes are avoided by taking the vertex and edges from the obstacles.

**Work flow of proposed OAGM model:**
- **Step 1:** Placing the obstacle in the simulation area i.e., Rectangle or Square
- **Step 2:** Placing the nodes randomly
- **Step 3:** Selecting the initial points of nodes and obstacle position are to be stored in files
- **Step 4:** Recursive function is used for selecting the source and destination nodes
- **Step 5:** Specify the limitations for speed and time
- **Step 6:** Find the Shortest path by using graph theory technique (Hybrid Bellman-Ford-Dijsktra Algorithm) until it reaches the destination
- **Step 7:** Check whether the obstacle is available if not reach the destination
- **Step 8:** If is an obstacle is available then step 4
- **Step 9:** Till Simulation time ends
- **Step 10:** Stop process

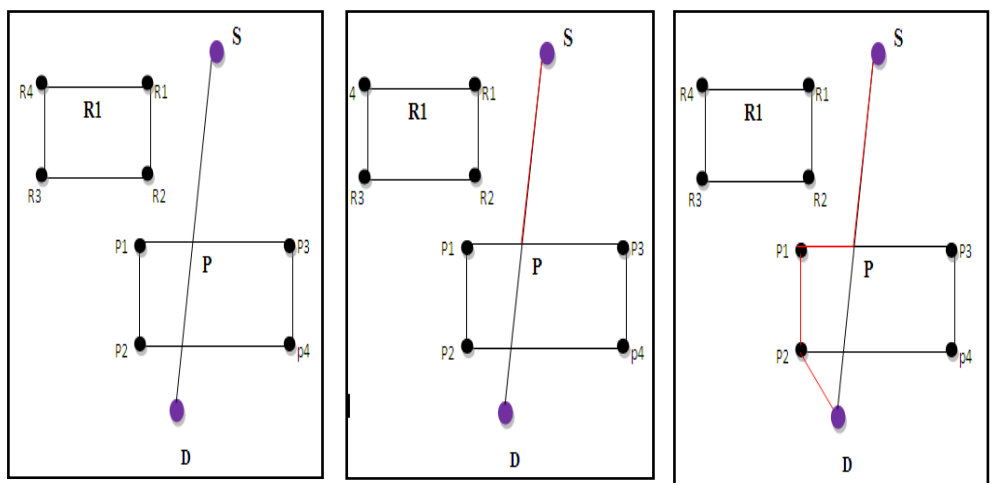

Fig. 1: The diagram shows how the model unobstructed line connection to Source and destination point

**Obstacle construction:** In OAGM model, rectangle or square shapes are used to specify the obstacles. In order to specify the obstacle the vertex, edges and coordinates are to be defined. The obstacles affect the node movements and the signals propagated through them. The nodes know the locations of all the obstacles that are fixed before simulation.

**Node movement process:** In the OAGM model, the initial placement of nodes and the destination points are selected with the nodes randomly based on a uniform distribution. Each node can move anywhere in the simulation terrain. Once the destination is determined, the node will compute the shortest path avoiding obstacles between the current and destination points by using recursive functions. Then, the node moves to the destination by following the shortest path with random speed and repeats until it reaches the new destination.

In this OAGM model, considers the corners and edges of the squares or rectangle. Although a certain things are verified in mathematical form and it is analyzed and no proof is obtained in the realistic mobility model. By this model we believe that it will provide a realistic balance with rectangle balance with a realistic mobility model which exists with random mobility models.

In Fig. 1 represents an example of how a node moves towards its destination point around the obstacles in the network area according to the OAGM model. The node U located in point S has set D at its destination point. The algorithm check is processed. If the direct line connecting S to D is unobstructed, realizing that obstacle P is in the way. From the vertices of the edge (P1, P3) that is the first one obstructing node U's way to the point D, the one closest to D is P1. Therefore, P1 is set as the next intermediate destination.

In order to reach P1, node U repeats the same process with S as its starting point and P1 as the destination. Obstacle Q obstructs node U to reach P1 immediately, so the closest to P1 edge p2 is selected. Then, the same process dictates that node U can move directly to S, since there are no obstacles between S and p1. Then it reaches the destination D.

In the proposed OAGM model, a destination point is set randomly, each node moves its way around the obstacles following a recursive process in order to reach it. If there is an unobstructed line of sight connecting the node with the destination point, the node follows this direct line to get to the desired destination. If there is an obstacle in the way, the node sets as its next intermediate destination the vertex or boundary of the visible obstacle that is closest to the destination and repeats the same process all over again with starting point its initial position and destination of the chosen vertex. We set undirected graph G (V, E) = Φ where V is a vertex set and E is an edge set. The computing shortest path from S to D is using Graph Theory Technique.

Once getting the Graph, we use the Hybrid Bellman-Ford-Dijkstra Algorithm to calculate the shortest paths between the destination point D and the other vertices of V. Thus, we get the shortest movement path from S to D which is shown in Fig. 1.

**Hybrid bellman-ford-dijkstra algorithm:** It is a combination of Bellman and Dijkstra algorithm. The model is a network of nodes connected by links. In this bellman-ford algorithm the delay time is calculated by using a transmitter. It finds a cheapest path from source S in a graph G with general edge costs. Dijsktra works for non-negative edges and also a search type algorithm. In this type of algorithm a single pass on all vertices and edges reachable from S.

**Algorithm:**
**Recall the basic Bellman-Ford (BF) and dijkstra algorithms:**
Initialization
d (v) ← ∞, for all v ϵ V
π (v) ← nil, for all v ϵ V
d (s) ← 0
Relax(u, v)
if d (u) + c (u, v) <d (v)
d (v) ← d(u) + c (u, v)
π (v) ← u
Plain scan
for each edge (u, v) E
Relax (u, v)
Dijkstra scan
S ← ϵ
while (there is a vertex in V\S with d<∞) do
find vertex u in V\S with the minimal value of d
S ← S ϵ {u}
for each edge (u, v) E/* scanning u*/
Relax (u, v)
Dijkstra (G, s)
Initialization
Dijkstra scan
return (d,)
Bellman-Ford (G, s)
Initialization
i ← 0
do
i++
Plain scan
until ((there was no change of d at Plain scan) or (i = |V|))
if (i<|V|) return (d, )
else return ("There exists a negative cycle reachable from s.")

Algorithm Bellman-Ford-Dijkstra (BFD) is as follows:

Bellman-Ford-Dijkstra (G, s)
Initialization
i ← 0
do
i++
Dijkstra scan
until ((there was no change of d at Dijkstra scan) or (i = |V| - 1))
if (i<|V| - 1) return (d,)
else return ("There exists a negative cycle reachable from s.")

Notice that BFD may be considered as a particular version of BF, since at each round, Relax is executed on all edges reachable from S (Dinitz and Itzhak, 2010).

**Lemma 1:** If there is no negative cycle reachable from s in G, BFD correctly computes the cheapest path value for all v V and the cheapest path tree in at most neg (G, s) + 2 rounds. Otherwise, it reports on the existence By proposition 2.2, if there is no negative cycle reachable from s,d values equal opt values at all vertices after neg (G, s) + 1 rounds of BFD. By Fact 2.1 (2), at the round numbered at most (neg (G, s) + 2), there is no change of d values and BFD stops. Since neg (G, s) ≤|V|-3, this should happen not later than after round |V|-1. Since Dijkstra scan executes Relax on all edges reachable from S, it may be considered a specific implementation of Plain scan. Therefore, BFD may be considered a specific implementation of the generic BF. Since BF is known to produce the cheapest path tree from s, this holds for BFD as well. If there exists a negative cycle reachable from s, then by Fact 2.1 (3), even at round |V|-1 there would exist an edge (u, v) reachable from s with d (u) +c (u, v) <d (v). Since Dijkstra scan executes Relax on all edges reachable from S, there would be a change of d at round |V|-1. BFD will then stop and report accordingly. The proof for BFD-p is similar. The above node movement mechanism resembles in a natural way how people behave and move their way around obstacles restricting their movement. When a person moves towards a destination point, it is rational to assume that they will try to go around the first observed obstacle in front and then over the vertex or boundary closest to the desired destination.

Even if eventually this will not be the most efficient choice in terms of the total distance covered until the destination is given that the overall obstacle placement is in most cases unknown, making this decision is the best a person can move into restricted area.

**Hierarchical node organization:**

- The nodes are organized in groups with a pre-defined leader/group.
- Group Size (i.e.,) each group contains certain no of nodes.
- Group Size is a parameter that can be act based on specific characteristics of the scenarios.
- Each member group is set the destination selection and a point within a constant distance from its leader's destination point referred as distance and begins towards it.

**Selection of source and destination:**

- The source and destination nodes are selected randomly from the total number of nodes simulated.
- We have taken approximately 5% of nodes in communication at any given time during the simulation interval.

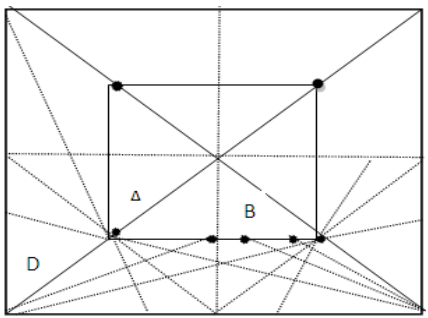

Fig. 2: Visualization of mobility scenarios and separation of area into sub regions

- Total 10% of the nodes will be either source or destination and remaining nodes will work as forwarding nodes.

**Signal propagation:** In order to make the integration of the obstacle completely natural, the propagation model in use to modify the effect of the physical layer phenomena likes evaporation, multi-path propagation, diffraction and reduction, etc., caused by the presence of obstacles is also taken into account. More specifically, the Two Ray Ground Propagation Model implemented in NS-2 is modified to ensure that when a signal is propagated through an obstacle, it suffers an attenuation randomly taken from a uniform distribution between fixed values. This model is actually a simplified version of propagation model used for Obstacle Mobility Model. The values used to test or verified are calculated from the range between 6 and 50 dB (Jardosh *et al.*, 2003; Christos *et al.*, 2012).

In this model the group mobility is followed. The separation of sub-regions of the simulation area is done using the following diagram. The movements are passed by the edge of the obstacle and follow the edge and boundary until it reaches the destination. In this OAGM, the general problem of finding the pdf fXY (x, y) of the node distribution for the OAGM model is calculated by unit square area with an obstacle of edge e located at its center, is reduced to calculating fXY (x, y) for the trapezoid ABCD as depicted in Fig. 2. Based on this area, fXY (x, y) in the remaining area can be obtained by replicating the results according to the axes of symmetry of the area (Fig. 3).

The trapezoid ABCD is further divided into nine sub-regions, based on the different shapes of the integrated parts. Each of the sub-regions R1 to R9 results in areas A (x, y, xs, ys, d) for the unobstructed movement and Qo (x, y, xd, yd) for the obstructed movement that have different shapes and thus require different calculations. Therefore, the fXY (x, y) is calculated for each of the sub regions R1 to R9 and the results are plotted symmetrically to the whole free space area.

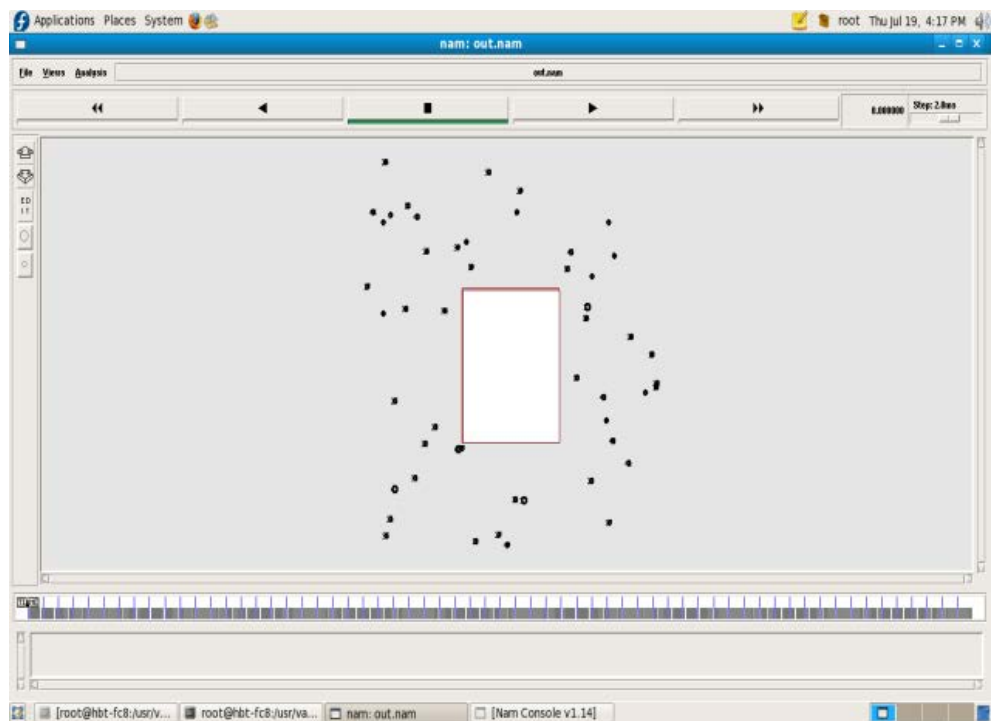

Fig. 3: Sample namfile of 50 nodes

## SIMULATION RESULTS

We implement the mobility model in NS2.30. The analysis of simulation results has been performed by means of the Trace Graph. The Studied Scenario of ad-hoc sensor network consists of 50 to 150 nodes with the interval of 50 nodes and with speed 0 to 10 ms intervals of 2 ms and the protocol is DSR with the parameters defined below in Table 1 and in Fig. 3 sample Nam file of 50 nodes with an obstacle in the center of the simulation terrain (1000×1000).

Mission Critical Model (MCM) with one obstacle in the center of the simulation area as Scenario 50, 100, 150, 200 and 250 nodes, respectively.

**Generated Packets (GP):** Here, the mobility models use the nodes 50-250 (with the interval nodes of 50) with different Speed 0 to 10 ms with the time interval of 2 ms (maximum speed = 10 m/s). The Generated Packets (GP) remains same even in the change of number of Speed varies (Table 2).

**Received Packets (RP):** It is defined as the number of packets received at the destination successfully. It is declared as Rf i.e., count of packets received from flow *f*.

In Table 3 shows the packets received at the destination. The proposed OAGM mobility models use the nodes 50-250 (with the interval nodes of 50) with different Speed 0 to 10 ms with the time interval of 2 ms (maximum speed = 10 m/s). The node 50, 100 and 150 the received packets are high when comparing generated packets to 200 and 250 nodes.

**Dropped Packets (DP):** The ratio of the data packets not delivered to the destinations to those generated from the sources. Mathematically, it can be expressed as:

$$DP = \frac{1}{N}\sum_{i=1}^{S}(ri + si)$$

where,
DP : The Number of Dropped Packets
i : Unique packet identifier
r*i* : Time at which a packet with unique id *i* is received
s*i* : Time at which a packet with unique id add with it
N : The number of connections, flows, *i* is sent

In Table 4 shows the drop the packets i.e., not received at the destination. The Proposed OAGM mobility models use the nodes 50-250 (with the interval nodes of 50) with different Speed 0 to 10 ms with the time interval of 2 ms (maximum speed = 10 m/s). The node 50, 100 and 150 the Dropped Packets are low when comparing generated packets to 200 and 250 nodes.

Table 1: Simulation parameter set

| | |
|---|---|
| Duration | 300s |
| Traffic sources | CBR, packet 512 byte, inter-arrival time-0.25s |
| Transport protocol | UDP |
| MAC protocol | Mac/802.11 |
| N/W interface | Phy/wireless phy |
| Propagation model | Two ray ground |
| Radius of node | 250 m |
| Antenna | Omni/antenna |
| Mobility models | MCM, proposed OAGM |
| No of nodes | 50-250 (interval nodes of 50) |
| Speed m/s | 0-10 m/s (interval speed of 2 m/s) |

Table 2: Generated packets

| 50 | 100 | 150 | 200 | 250 |
|---|---|---|---|---|
| 3480 | 5798 | 9272 | 11586 | 13898 |

Table 3: Received Packets (RP) in terms of packets

| No of nodes | Speed 2 m/s | Speed 4 m/s | Speed 6 m/s | Speed 8 m/s | Speed 10 m/s |
|---|---|---|---|---|---|
| 50 | 3480 | 3445 | 3427 | 3410 | 3440 |
| 100 | 5790 | 5755 | 5735 | 5650 | 5686 |
| 150 | 9212 | 8700 | 7820 | 9100 | 962 |
| 200 | 6700 | 4012 | 5070 | 661 | 318 |
| 250 | 1326 | 639 | 642 | 269 | 120 |

Table 4: Dropped Packets (DP) in terms of packets

| No of nodes | Speed 2 m/s | Speed 4 m/s | Speed 6 m/s | Speed 8 m/s | Speed 10 m/s |
|---|---|---|---|---|---|
| 50 | 0 | 35 | 53 | 70 | 40 |
| 100 | 8 | 43 | 63 | 148 | 112 |
| 150 | 60 | 572 | 1452 | 172 | 8310 |
| 200 | 4886 | 7574 | 6516 | 10925 | 11268 |
| 250 | 12572 | 13259 | 13256 | 13629 | 13778 |

Table 5: Control Overhead (CO) in terms of packets

| No of nodes | Speed 2 m/s | Speed 4 m/s | Speed 6 m/s | Speed 8 m/s | Speed 10 m/s |
|---|---|---|---|---|---|
| 50 | 105 | 1058 | 2015 | 1867 | 2001 |
| 100 | 555 | 2164 | 3377 | 4990 | 3946 |
| 150 | 2660 | 13888 | 19878 | 7320 | 280861 |
| 200 | 63436 | 229250 | 162985 | 590990 | 648099 |
| 250 | 563427 | 687386 | 717255 | 874759 | 1039263 |

**Control Overhead (CO):** The control overhead is defined as the total number of control packets (i.e., N*f*) exchanged successfully.

In Table 5 shows the Packets Exchanged successfully. The Proposed OAGM mobility model and MCM model giving the same result by using the nodes 50-250 (with the interval nodes of 50) with different Speed 0 to 10 ms with the time interval of 2 ms (maximum speed = 10 m/s). The node 50, 100 and 150 the Control overhead are low when comparing generated packets to 200 and 250 nodes.

**Packet Delivery Ratio (PDR):** The ratio of the data packets delivered to the destinations to those generated from the sources. Mathematically, it can be expressed as:

$$p = \frac{1}{c}\sum_{f=1}^{s}\frac{Rf}{Nf} * 100$$

Table 6: Packet Delivery Ratio (PDR) in terms of percentage (%)

| No of nodes | Speed 2 m/s | Speed 4 m/s | Speed 6 m/s | Speed 8 m/s | Speed 10 m/s |
|---|---|---|---|---|---|
| 50 | 99.9425 | 98.3046 | 97.7874 | 97.5000 | 98.2759 |
| 100 | 99.8103 | 98.9997 | 98.7237 | 97.0852 | 97.8441 |
| 150 | 99.2990 | 93.5828 | 84.2150 | 97.9770 | 10.2351 |
| 200 | 60.2106 | 34.4813 | 43.5526 | 5.57569 | 2.64112 |
| 250 | 15.2166 | 4.48266 | 4.49705 | 1.8276 | 0.7627 |

Table 7: Average end to End Delay (ED) in terms of milli seconds (m/s)

| No of nodes | Speed 2 m/s | Speed 4 m/s | Speed 6 m/s | Speed 8 m/s | Speed 10 m/s |
|---|---|---|---|---|---|
| 50 | 16.5900 | 19.6230 | 60.9207 | 72.4227 | 31.0448 |
| 100 | 22.7325 | 17.9425 | 44.0934 | 23.5896 | 39.7311 |
| 150 | 47.2069 | 230.4370 | 72.9717 | 52.3945 | 131.1000 |
| 200 | 572.6870 | 120.9920 | 35.4563 | 226.0120 | 101.1170 |
| 250 | 96.1891 | 107.4080 | 72.2491 | 180.1030 | 111.0600 |

where,
p  = The Ratio of successfully delivered packets
C  = The total number of flow or connections
f  = The unique flow id serving as an index
R$f$ = The count of packets received from flow $f$
N$f$ = The count of packets transmitted to $f$

In Table 6 shows the percentage of packets delivered to the destination successfully. The proposed OAGM mobility models use the nodes 50-250 (with the interval nodes of 50) with different Speed 0 to 10 ms with the time interval of 2 ms (maximum speed = 10 m/s). The node 50, 100 and 150 the delivery ratio is high when compared to 200 and 250 nodes.

**Average end to End Delay (ED):** This includes all possible delays caused by buffering during route discovery latency, queuing at the interface queue, retransmission delays at the MAC and propagation and transfer times. It can be defined as:

$$D = \frac{1}{N}\sum_{i=1}^{S}(ri - si)$$

where,
D  = The number of successfully received packets
$i$  = Unique packet identifier
$ri$ = Time at which a packet with unique id $i$ is received
$si$ = Time at which a packet with unique id $i$ is sent
D  = Measured in m/s. It should be less for higher performance

In Table 7 shows the Average End to End Delay. The Proposed OAGM mobility models use the nodes 50-250 (with the interval nodes of 50) with different Speed 0 to 10 ms with the time interval of 2 ms (maximum speed = 10 m/s). The node 50, 100 and 150 the delay time is low when compared to 200 and 250 nodes.

**Performance analysis of proposed OAGM and MCM model:** In Table 8 shows the Packet delivery Ratio. The proposed OAGM and MCM mobility models use the nodes 50-250 (with the interval nodes of 50) with different Speed 0 to 10 ms with the time interval of 2 ms (maximum speed = 10 m/s). The Proposed OAGM model gives better result than the MCM model.

In Table 9 shows the Average End to End Delay. The proposed OAGM and MCM mobility models use the nodes 50-250 (with the interval nodes of

Table 8: Packet Delivery Ratio (PDR) in term of percentage (%)

| Models/speed | 2 m/sec | 4 m/sec | 6 m/sec | 8 m/sec | 10 m/sec |
|---|---|---|---|---|---|
| MCM (50 node) | 99.94 | 98.30 | 97.79 | 97.50 | 98.28 |
| Proposed OAGM (50 node) | 100 | 98.99 | 98.48 | 97.99 | 98.85 |
| MCM (100 node) | 99.81 | 98.99 | 98.72 | 97.08 | 97.84 |
| Proposed OAGM (100 node) | 99.86 | 99.26 | 98.91 | 97.45 | 98.06 |
| MCM (150 node) | 99.30 | 93.58 | 84.22 | 97.98 | 10.24 |
| Proposed OAGM (150 node) | 99.35 | 93.83 | 84.34 | 98.14 | 10.37 |
| MCM (200 node) | 60.21 | 34.48 | 43.55 | 5.57 | 2.64 |
| Proposed OAGM (200 node) | 57.83 | 34.63 | 43.76 | 5. 71 | 2.75 |
| MCM (250 node) | 15.21 | 4.48 | 4.49 | 1.83 | 0.76 |
| Proposed OAGM (250 node) | 9.54 | 4.59 | 4.62 | 1.94 | 0.86 |

Table 9: Avg end to end delay in term of milli seconds

| Models/speed | 2 m/sec | 4 m/sec | 6 m/sec | 8 m/sec | 10 m/sec |
|---|---|---|---|---|---|
| MCM (50 node) | 17.06 | 20.62 | 65.92 | 76.42 | 34.04 |
| Proposed OAGM (50 node) | 16.59 | 19.62 | 60.92 | 72.42 | 31.04 |
| MCM (100 node) | 26.63 | 20.94 | 49.09 | 28.59 | 42.53 |
| Proposed OAGM (100 node) | 22.73 | 17.94 | 44.09 | 23.59 | 39.73 |
| MCM (150 node) | 50.41 | 237.33 | 77.67 | 58.29 | 138.80 |
| Proposed OAGM (150 node) | 47.21 | 230.43 | 72.97 | 52.39 | 131.10 |
| MCM (200 node) | 576.89 | 124.84 | 40.06 | 230.31 | 105.72 |
| Proposed OAGM (200 node) | 572.68 | 120.99 | 35.46 | 226.01 | 101.11 |
| MCM (250 node) | 345.54 | 111.50 | 78.72 | 184.90 | 115.26 |
| Proposed OAGM (250 node) | 96.19 | 107.41 | 72.25 | 180.10 | 111.06 |

50) with different Speed 0 to 10 ms with the time interval of 2 ms (maximum speed = 10 m/s). The proposed OAGM model gives better result than the MCM model.

## CONCLUSION

The main aim is to study the MCM Realistic Mobility Model and proposed OAGM model is to prove that the OAGM mobility model outperforms well. Mobility model extremely affects the performance results of a routing protocol in a realistic environment. The main characteristics of the proposed OAGM model with which the real-life properties of movement in such environments are captured by the presence of an obstacle which affects the node movement and signal propagation, source and destination selection and various node movements based on graph theory technique followed by the nodes. To the best of our knowledge, the MCM is the first model on Realistic movement and the Second is the proposed Obstruction Avoidance Generously Mobility model (OAGM) which deals with all these issues in such a complete way. By analyzing the MCM and OAGM model, the overall performance of this proposed OAGM model is 2% better than the MCM model.

## ACKNOWLEDGMENT

I thank the Karpagam University for the motivation and encouragement to do research work in a successful way.